\documentclass[twocolumn,superscriptaddress,pre]{revtex4}
\usepackage{physics}
\usepackage{hyperref}
\usepackage{xcolor,graphicx}
\usepackage{subfigure}
\usepackage{amssymb}
\usepackage{natbib}
\usepackage{amsmath}
\usepackage{amsfonts}
\usepackage{mathrsfs}
\usepackage{dcolumn} 
\usepackage{bm}
\usepackage{cancel}
\usepackage{mathbbol}
\usepackage{epsfig}
\usepackage{units}
\usepackage{esint}
\usepackage{soul} 
\usepackage{url}
\newcommand{\beq}{\begin{equation}}
\newcommand{\eeq}{\end{equation}}

\begin{document}

\title[]{New Trends in Quantum Machine Learning}

\author{Lorenzo Buffoni}
\address{Dipartimento di Fisica e Astronomia, Universit\'{a} di Firenze, I-50019 Sesto Fiorentino, Italy}
\address{Dipartimento di Ingegneria dell'Informazione, Universit\'{a} di Firenze, I-50139 Firenze, Italy}
\author{Filippo Caruso}
\address{Dipartimento di Fisica e Astronomia, Universit\'{a} di Firenze, I-50019 Sesto Fiorentino, Italy}
\address{LENS, QSTAR and CNR-INO, I-50019 Sesto Fiorentino, Italy}

\begin{abstract}
Here we will give a perspective on new possible interplays between Machine Learning and Quantum Physics, including also practical cases and applications. We will explore the ways in which machine learning could benefit from new quantum technologies and algorithms to find new ways to speed up their computations by breakthroughs in physical hardware, as well as to improve existing models or devise new learning schemes in the quantum domain.
Moreover, there are lots of experiments in quantum physics that do generate incredible amounts of data and machine learning would be a great tool to analyze those and make predictions, or even control the experiment itself. On top of that, data visualization techniques and other schemes borrowed from machine learning can be of great use to theoreticians to have better intuition on the structure of complex manifolds or to make predictions on theoretical models. This new research field, named as Quantum Machine Learning, is very rapidly growing since it is expected to provide huge advantages over its classical counterpart and deeper investigations are timely needed since they can be already tested on the already commercially available quantum machines.
\end{abstract}

\maketitle

\section{Introduction}

Machine learning (ML) \cite{bishop_pattern_2011,cover_elements_1991,hastie2009elements,hundred} is a broad field of study, with multifaceted applications of  cross-disciplinary  breadth. ML ultimately aims at developing computer algorithms that improve automatically through experience. The core idea of artificial intelligence (AI)  technology is that systems can learn from data, so as to identify distinctive patterns and make consequently decisions, with minimal human intervention. The range of applications of ML methodologies is extremely vast \cite{sutton2018reinforcement,graves2013speech,sebe2005machine,grigorescu2020survey}, and still growing at a steady pace due to the pressing need to cope with the efficiently handling of big data \cite{chen2014big}. 
Training and deployment of large-scale machine learning models faces computational challenges \cite{tieleman2008training} that are only partially met by the development of special purpose classical computing units such as GPUs. This has led to an interest in applying quantum computing to machine learning tasks \cite{schuld2015introduction,wittek2014quantum,adcock2015advances,arunachalam2017survey,Biamonte:2017db} and to the development of several quantum algorithms \cite{harrow2009quantum,wiebe2012quantum,childs2015quantum,lloyd2014quantum} with the potential to accelerate training. Most quantum machine learning algorithms need fault-tolerant quantum computation \cite{nielsen2002quantum,lidar2013quantum,fowler2012surface}, which requires the large-scale integration of millions of qubits and is still not available today. It is however possible that quantum machine learning (QML) will provide the first breakthrough algorithms to be implemented on commercially available Noisy Intermediate Scale Quantum (NISQ) devices \cite{farhi2001quantum,johnson2011quantum,neill2017blueprint,kandala2017hardware}. Indeed, a number of interesting breaktrhoughs have alreasy been made at the interface of Quantum Physics and Machine Learning \cite{carleo2019machine}. For example in many-body quantum physics Machine Learning has been successfully employed to speed up simulations \cite{arsenault2017projected}, predict phases of matter \cite{carrasquilla2017machine} or find variational ansatz for many body quantum states \cite{carleo2017solving}. Similarly in quantum computation Machine Learning has been recently found success in quantum control \cite{bukov2018reinforcement} and to provide error correction \cite{varsamopoulos2020decoding}. \\
In this paper we will give our perspective on new possible trends of machine learning in the quantum domain, covering all the main learning paradigms and listing some very recent results. Thus we will try to briefly define the scope of each one of these ML domains, referring to the literature for a complete overview of the specific topics. The three main classes of learning algorithms are the following ones:
\begin{itemize}
    \item \textbf{Supervised learning}: In supervised learning \cite{bishop_pattern_2011, goodfellow2016deep} one deals with an annotated dataset $\left\lbrace(x_i,y_i) \right\rbrace_{i=1}^N$. Each element $x_i$ is called an input or feature vector. It can be the vector of pixel values of an image or a feature such as height, weight and gender and so on. All input data $x_i$ of the same dataset share the same features (with different values). The label $y_i$ it is the ground truth upon which we build the knowledge of our learning algorithm. It can be a discrete class in a set of possible objects or a real number representing some property we want to predict, or even some complex data structure. For example if we want to build a spam classifier the labels will be $y_i=1$ (spam) or $y_i=0$ (not spam). The goal of a supervised learning algorithm is to use the dataset to produce a model that, given an input vector $x$, can predict the correct label $y$.
    
    \item \textbf{Unsupervised learning}: In unsupervised learning \cite{vincent2008extracting,hinton1995wake,bengio2007greedy} the dataset is a collection of unlabeled vectors $\left\lbrace x_i \right\rbrace_{i=1}^N$. The goal of unsupervised learning is to take this input vector and extract some useful property from the single data or the overall data distribution of the dataset. Examples of unsupervised learning are: clustering where the predicted property is the cluster assignment; dimensionality reduction where the distribution of data is mapped in a lower dimensional manifold; outlier detection where the property predicted is the 'typicality' of the data with respect to its distribution; and generative models where we want to learn to generate new points from the same distribution of the dataset.
    
    \item \textbf{Reinforcement learning}: The ML subfield of reinforcement learning \cite{sutton2018reinforcement} assumes that the machine 'lives' in an environment and can probe the state of the environment as a feature vector. The machine can perform different actions at different states bringing to different rewards. The goal of this machine (or agent) is to learn a policy/strategy. A policy is a function that associates to a particular feature vector, representing the state of the environment, the best action to execute. The optimal policy maximizes the expected average reward. Reinforcement learning has been widely employed in scenarios where decision making and long-term goals are crucial, for example in playing chess, controlling robots or logistics in complex environments.
\end{itemize}

\begin{figure}
    \centering
    \includegraphics[width=0.9\linewidth]{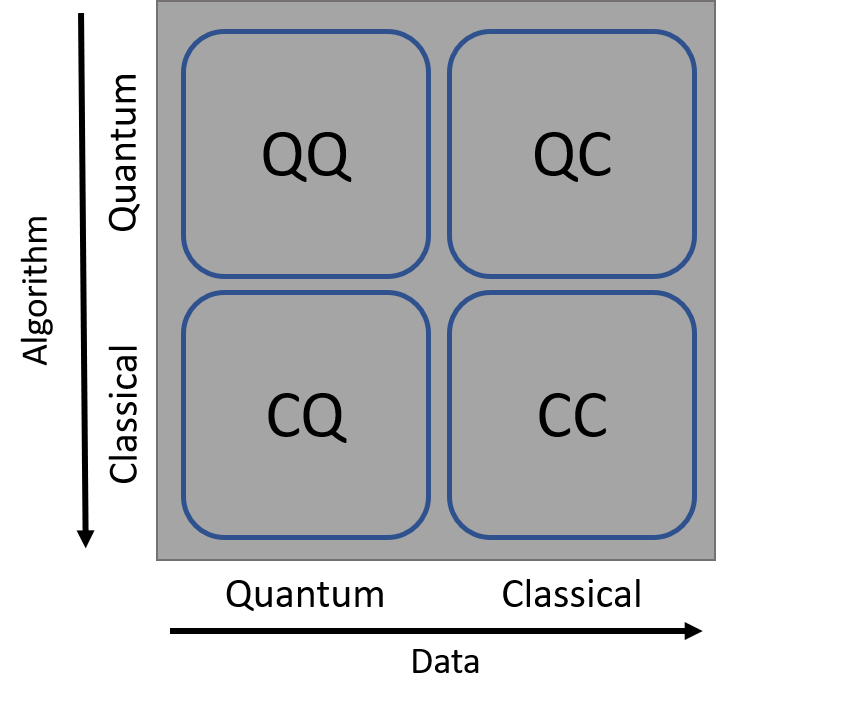}
    \caption{Scheme of the various types of QML. The data to be processed can be either classical or quantum and the algorithm that process the data itself can also be either classical or quantum. That gives four possible combinations of algorithms, three of which (QQ, QC, CQ) usually fall under the umbrella of QML.}
    \label{fig:qml_scheme}
\end{figure}

Now it is important to point out that there is no rigorous universal definition of QML and instead it has assumed different meanings in the context where it has been studied in literature. The main subjects are the data that can be either classical or quantum, and the algorithm used to perform ML on the data itself, which can also be classical or quantum. One indeed can define four macro categories of algorithms \cite{Dunjko2016} as depicted in Fig.\ref{fig:qml_scheme}, three of which are generally considered QML. There are also a lot of gray areas where the algorithms are hybrid quantum-classical or only a subroutine or an optimization task is carried out by the quantum processor. An exhaustive list of all possible interplays between quantum and classical ML algorithms is outside the scope of this paper, so we will stick to this simplified representation for the sake of clarity.

\section{Supervised Learning}

\begin{figure}
\centering
\subfigure[]{\includegraphics[width=.6\linewidth]{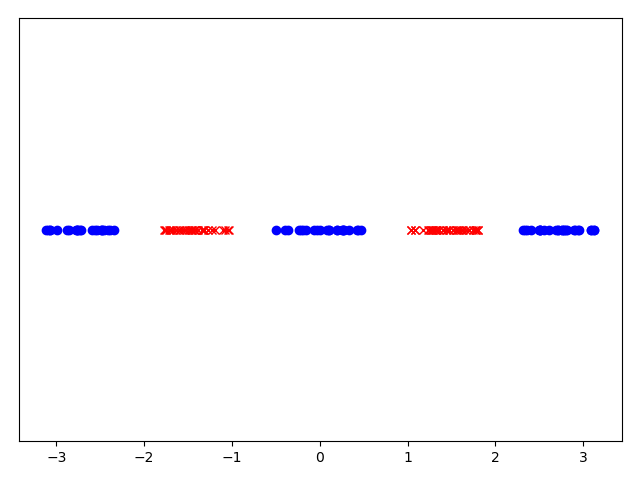}}
\subfigure[]{\includegraphics[width=.99\linewidth]{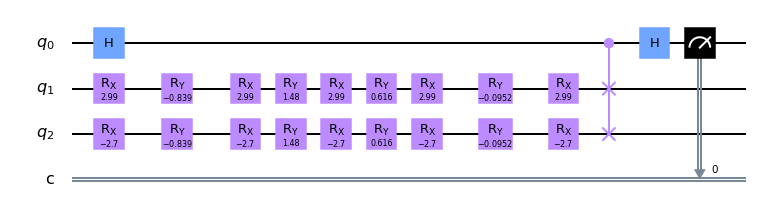}}
\caption{In panel (a) the 1D syntetic dataset used as classification benchmark. Class A are the blue dots while class B are the red crosses. Note that in the 1D space the dataset cannot be linearly classified, i.e. there is not a simple threshold allowing us to separate the two classes. In panel (b) an example of the embedding circuit including the final SWAP test to compute the overlap between the two embeddings \cite{embedding2020preparation}.}
\label{fig:circuit}
\end{figure}

There are already in literature several examples of supervised learning applications for NISQ devices. For instance, small gate-model devices and quantum annealers have been used to perform quantum heuristic optimization \cite{kadowaki1998quantum,santoro2002theory,brooke2001tunable,farhi2014quantum,peruzzo2014variational,kandala2017hardware} and to solve classification problems \cite{neven2008training,denchev2012robust,pudenz2013quantum,mott2017solving}. An interesting approach to classification was devised in Ref.  \cite{lloyd20}, where classical data are embedded into a larger quantum (Hilbert) space describing the state of a quantum system. The idea, which is similar in spirit to classical Support Vector Machines (SVMs) \cite{bishop_pattern_2011}, is to map classical data into an high dimensional space where the classes can be well separated by an hyperplane. In the quantum case the embedding is done by a quantum circuit composed of single- and multi-qubit gates, effectively mapping the classical data into the Hilbert space of the qubits. For example, an embedding of a classical point $x$ into a single qubit state $\ket{x}$ could be the following:
\begin{equation}
    R_X(x)R_Y(\theta_3)R_X(x)R_Y(\theta_2)R_X(x)R_Y(\theta_1)R_X(x)\ket{0} \longrightarrow \ket{x}.
\end{equation}
where $R_X$ and $R_Y$ are rotation operator along the axes $X$ and $Y$, respectively, while the rotation angles $\lbrace \theta_1, \theta_2, \theta_3 \rbrace$ are the trainable parameters of our learning model. Once a dataset $\lbrace x_i \rbrace_{i=1}^N$ is embedded, one can compute (using a SWAP test) the overlaps $|\braket{x_i}{x_j}|^2$. Points (states) belonging to the same class will have an overlap close to 1 while points from different classes will have an overlap close to 0, hence enabling the classification of the dataset. The training can be done using software such as Pennylane \cite{pennylane} that computes the gradients of the parametric quantum gates. In Fig. \ref{fig:circuit} we show an embedding circuit for an one-dimensional dataset that can be trained by small quantum processors with the circuit proposed above.
\begin{figure}
\centering
\subfigure[]{\includegraphics[width=.49\linewidth]{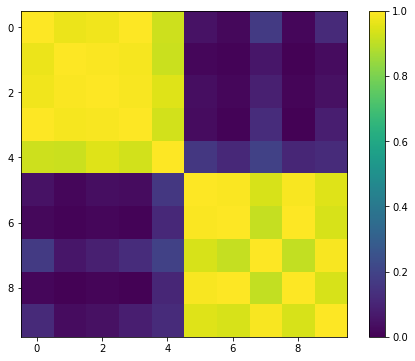}}
\subfigure[]{\includegraphics[width=.49\linewidth]{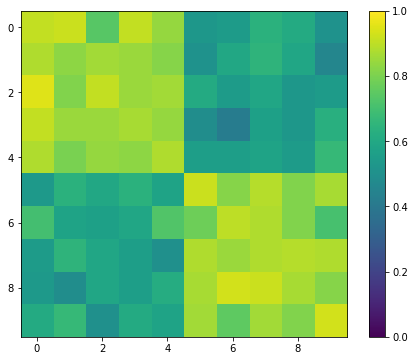}}
\caption{(a) Theoretical Gram matrix, representing the overlaps between 10 test embeddings; (b) Experimental Gram matrix on the IBM Valencia QPU. The achieved results are pretty consistent with the theory albeit a bit noisier than other setups that are exploited in Ref. \cite{embedding2020preparation}.}
\label{fig:ibm}
\end{figure}

In Ref. \cite{embedding2020preparation} we train an embedding of that dataset and test it with $10$ validation points on a real quantum processor. The process has been carried out in the IBM quantum platform using the Valencia QPU (Quantum Processing Unit) composed of 5 qubits. The number of samples was rather small, namely $100$ samples per point. That is due to the queues to access the system and to the fact that we need to compute $100$ overlaps to build the Gram matrix for our $10$ test points. As one can see in Fig. \ref{fig:ibm} our results \cite{embedding2020preparation}, even if they contain some noise, are good and able to achieve a good classification boundary between the two classes. That is quite remarkable as this scheme is implemented on a real NISQ device via the cloud. Let us also point out that the error on the overlap depends on the number of samples we take and it could be the case that our $100$ samples do not contain enough statistics. This method has then been proved effective on small datasets (embedded in 1 or 2 qubits) and should be quite robust to experimental noise, yielding some promises for the future of NISQ devices.
Indeed, as argued in Ref. \cite{lloyd20}, we can map high-dimensional data in a \textit{n-qubit} state using the same protocol. And, if we could build a circuit of $100$ qubits with circuit depth $100$ and a decoherence time of $10^{-3} s$, it could be capable to embed $\mathcal{O}(10^{10})$ bits of classical information, i.e. a task which is classically unattainable. Therefore, high-dimensional embeddings of large data sets for (QC-type) QML could be accessible in future on NISQ devices.

Once classical data are embedded on quantum states, one has to deal with a fundamental problem in quantum physics that is the discrimination amongst a set of non-orthogonal quantum states of a system. This can be addressed for instance by following a recent approach \cite{dallapozza2020} based on the quantum stochastic walk model \cite{whitfield2010, Caruso2014} and inspired by the structure of classical neural networks. In particular, the quantum states to discriminate can be encoded on the input nodes of a network, while the discrimination is achieved on the output nodes. Then, the parameters of the underlying network (e.g. hopping rates or link weights) are optimized to obtain the highest probability of correct discrimination. Interestingly enough, this probability approaches the theoretical optimal quantum limit known as \emph{Helstrom bound} \cite{Helstrom1976}. Our numerical and analytical results in Ref.
\cite{dallapozza2020} also show the robustness and reconfigurability of the network for different set of quantum states. Our proposed (QQ-type) architecture can pave the way for experimental realizations of QML protocols as well as novel quantum generalizations of deep learning algorithms running over larger complex networks.

Finally, let us mention a CQ-type QML application we are proposing in Ref. \cite{martina2020} where artificial neural network models \cite{Goodfellow-et-al-2016,bishop_pattern_2011} are exploited for quantum noise discrimination in stochastic quantum dynamics. In particular some Hamiltonian parameters are affected by noise, which is described as a stochastic process associated to a specific probability distribution. Our aim is to discriminate among different noise probability distributions and correlation parameters, by measuring the quantum state at discrete time instants. In particular, we proposed the use of classical ML algorithms such as SVMs and Recurrent Neural Networks (RNN), i.e. Long Short Term Memory (LSTM)\,\cite{cho_gru2014} and Gated Recurrent Unit (GRU)\,\cite{gers_lstm1999}, to perform supervised classification tasks on simulated quantum data. We believe that these are preliminary steps towards QML algorithms for quantum sensing \cite{DegenRMP2017}, e.g. to detect Markovian vs. non-Markovian noise in real quantum systems as optical and atomic platforms that are very promising candidates for NISQ devices.


\section{Unsupervised Learning}

Unlike supervised learning \cite{goodfellow2016deep}, unsupervised learning is a much harder, and still largely unsolved, problem. And yet, it has the appealing potential to learn the hidden statistical correlations of large unlabeled datasets \cite{vincent2008extracting,hinton1995wake,bengio2007greedy}, which constitute the vast majority of data being available today. 

For instance, clustering algorithms have the ability to separate unlabeled data in different classes (or clusters) without the need for external labeling and supervision. Hybrid approaches have been tried to solve clustering \cite{otterbach2017unsupervised} by applying a quantum-based optimization scheme to a classical clustering algorithm. In order to represent the clusters, we need to define a distance measure $d(x_i,x_j)$  between two data elements $x_i$ and $x_j$. For example, a choice could be the Euclidean distance, but specific applications may naturally lead to very different metrics. You can then calculate the distance matrix $C$ with elements $C_{ij}=d(x_i,x_j)$. Now this matrix can also be interpreted as adjacency matrix of a weighted graph $G$ and, by a suitable choice of metric and doing some coarse graining, the problem of clustering reduces to the MAXCUT optimization problem \cite{mahajan1999parameterizing} on the graph $G$. The MAXCUT problem  is  an  example  of  the  class of  NP-complete problems,  which  are  notoriously hard to be solved. In Ref. \cite{otterbach2017unsupervised} an hybrid scheme is devised where the MAXCUT optimization is solved by the QAOA algorithm \cite{farhi2014quantum} and applied to a syntetic dataset composed of 2 clusters of 10 points living in a 2-dimensional space. The model was also successfully tested on a real 19-qubit chip produced by Rigetti as demonstration of a QC-type application.

Another opportunity to use quantum technologies on unsupervised learning tasks comes from Variational Autoencoders (VAEs) \cite{kingma2014semi}. Variational Autoencoders are generative models in which the goal is to learn a complex data distribution (e.g. the pixels of images representing cats) to later sample from it 'generating' new objects. In particular VAEs learn a probability distribution $p_{\theta}(x,z) = p_{\theta}(x|z)p_{\theta}(z)$, where usually the a {\it posterior} distribution $p_{\theta}(x|z)$ is implemented by a Deep Neural Network and $p_{\theta}(z)$ is simple (prior) distribution (e.g., i.i.d Gaussian or Bernoulli variables). The idea here is to replace the distribution $p_{\theta}(z)$ with a complex distribution sampled from a quantum device \cite{khoshaman2018quantum,vinci2020path}. This setup allows for the construction of quantum-classical hybrid generative models that can be scaled to large, realistic datasets. For example in Ref. \cite{vinci2020path} a quantum-classical hybrid VAE was trained using a D-Wave 2000Q quantum annealer on the popular MNIST \cite{lecun1998mnist} dataset of handwritten digits.
This quantum-classical hybrid VAE still employs a large amount of classical computing power performed on modern GPUs. The computational task that we offloaded to the quantum annealer (sampling from a complex distribution) can still be performed classically at a fraction of the overall computational cost. To achieve any form of quantum advantage in this framework, we need to offload generative capacity to the prior distribution, by exploiting large graphs capable of representing complex probability distributions from which classical sampling becomes too expensive. We have evidence that this path to quantum advantage is possible by deploying annealers with denser connectivities and lower noise, engineering classical neural networks that better exploit physical connectivities and by working with more complex datasets. All these improvements should be achievable in the near future, and represent possible interesting lines of research in the area of QC-type QML models.

Furthermore, one of the most remarkable results in unsupervised ML is provided by generative adversarial networks (GANs) \cite{goodfellow2014generative,goodfellow2016nips}, where game-theory models aim to learn how to faithfully reproduce some given distribution of data. In particular, the idea is that two agents, named as the generator and the discriminator, compete against each other in a zero-sum game where on one side the generator aims to generate 'fake' data that the discriminator is not able to distinguish from the real generated ones, while on the other side the latter optimizes its discrimination strategy. Under some reasonable assumptions, it is possible to demonstrate that the game reaches the unique (Nash) equilibrium point where the generator is able to exactly reproduce the desired (real) data distribution. GANs have been generalized to the quantum case leading to the so-called Quantum Generative Adversarial Networks (QGANs) \cite{lloyd2018quantum,dallaire2018quantum}. They are an example of QQ-type QML. Again the goal is to learn of reproducing the state of a physical system that is now quantum, e.g. a register of qubits. This can be for instance implemented by exploiting parametrized quantum circuits (where quantum gates are controlled by real tunable parameters) \cite{benedetti2019parameterized} allowing to realize, among others, quantum approximation optimization algorithms  \cite{farhi2014quantum}, VAEs \cite{pepper2019experimental}, and eigensolvers \cite{peruzzo2014variational}. However, recent efforts have been mainly focused on learning pure states \cite{dallaire2018quantum,benedetti2019adversarial}, while the scenario of mixed quantum states is currently under investigation and some preliminary results are appearing in literature \cite{hu2019quantum,braccia2020}. In particular, in Ref. \cite{braccia2020} we show how the emergence of limit cycles may considerably extend the convergence time in the case of mixed quantum (generated data) states that of course play a key role for practical implementations on commercially available NISQ devices.


\section{Reinforcement Learning}

Reinforcement learning has been already applied to closed quantum systems, i.e. following a unitary dynamical evolution \cite{Dunjko2016}. Yet, the setting of an agent acting on an environment has a natural analogue in open quantum systems~\cite{Breuer2002}.
In Ref. \cite{maze2020preparation} we propose to generalize RL into the quantum domain in order to solve a Quantum Maze problem. It is an example of QQ-type QML. For Quantum Maze we mean a network whose topology is represented by a perfect maze, i.e. there is always one unique path between any two points in the maze, and the dynamical evolution of the physical state (described by its adjacency matrix $\rho$) of this network is quantum. The paths of the maze are defined by links between pair of nodes, which are described by an adjacency matrix $A$. The coefficient $A_{i,j}=1$ indicates the presence of the link, while $A_{i,j}=0$ indicates its absence. 
The evolution of the quantum system has been based on the quantum stochastic walk model \cite{whitfield2010, Caruso2014} that can be described by a Lindblad master equation \cite{lindblad1976generators} where the Hamiltonian of the system $H$ corresponds to the adjacency matrix itself, i.e. $H=A$, and with a set of Lindblad operators acting as noise for the quantum walker, i.e.
\beq
    {\dot \rho} = -(1-p)\ i[H,\rho] + p\  \mathcal{L}_{CRW} (\rho) + \mathcal{L}_{sink} (\rho)
    \label{densityEvolution}
\eeq
with a term describing (for $p=1$) the regime of a classical random walk (CRW), as
\beq
    \mathcal{L}_{CRW} (\rho) = \sum_{i,j} L_{ij} \rho L_{ij}^{\dagger} - \frac{1}{2} \{L_{ij}^\dagger L_{ij}, \rho\},
\eeq
with the Lindblad operators being defined as $L_{ij} = (A_{ij}/d_j)\ket{i}\bra{j}$, where $\lbrace d_j \rbrace$ are the node degrees. When $p=0$ one recovers the pure quantum walker regime, while $p=1$ corresponds to the classical scenario for a random walker. For the intermediate values of $p$ in the range $]0,1[$, one has a quantum stochastic walker with an interplay between coherent evolution and noise effects. In addition, there is a Lindblad operator that irreversibly transfers the population from the ``exit" node $n$ to the sink $S$ with a rate $\Gamma$, as follows
\beq
    \mathcal{L}_{sink} (\rho) = \Gamma \left [ 2\ket{S}\bra{n}\rho \ket{n} \bra{S} - \{\ket{n}\bra{n}, \rho\} \right ].
\eeq
Moreover, we assume that all population is initially at the entrance node of the maze. The exit of the maze corresponds to the sink $S$ which irreversibly traps the population. Equation \ref{densityEvolution} gives leads to the following expression for the escaping probability from the maze 
\beq
    p_{sink}(t) = 2 \Gamma \int_0^t \rho_{n,n}(t') \ \dd t' \; .
\eeq
The goal is to maximize this escaping probability in the shortest amount of time. Since we consider a perfect maze, there is a single path to exit the maze from the entrance node, in presence of several dead ends.  
\begin{figure}[ht]
    \centering
    \subfigure[]{\includegraphics[width=0.7\linewidth]{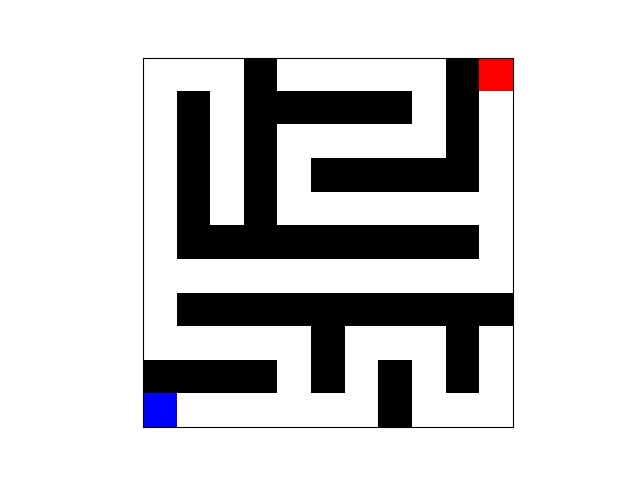}}
    \subfigure[]{\includegraphics[width=0.99\linewidth]{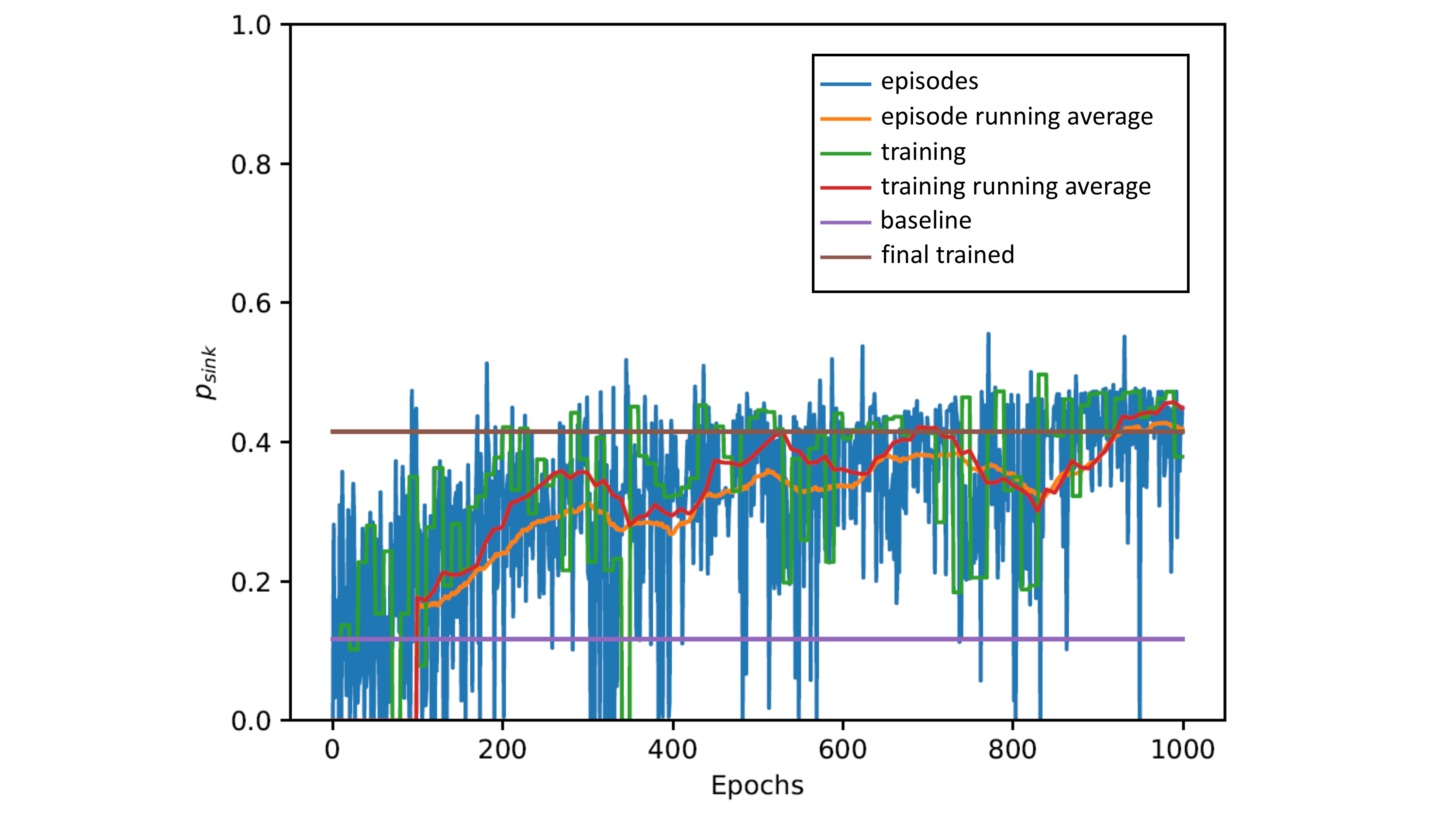}}
    \caption{ In panel (a) a sample of a $6 \times 6$ perfect maze. In white the possible paths, in black the walls. In the lower left corner, the node in blue is the entrance to the maze, corresponding to the initial quantum state, while the node in red is the exit, i.e., the node connected to the sink. In panel (b) an example of a training curve for an agent doing actions at periodic time instants, in the regime of $p=0.8$ on the $6 \times 6$ maze in (a). The curves show the rewards from the single episodes and their running average over 100 episodes as well as the training curve of the agent with the running average again over 100 episodes. The two constant lines are the baseline quantum walker with no actions performed by the agent and the final trained policy. \cite{maze2020preparation}}
    \label{fig:training_curve}
\end{figure}
In this scenario, an external controller (the computer user or the quantum system itself) is the agent that has some information about the quantum state of the system, while the maze is the RL environment. The available actions for the agent, during the evolution, are:
\begin{itemize}
    \item[1 --] Building walls. During the evolution, at given (periodic) time instants, the system can change the adjacency matrix of the environment removing a link (changing from $1$ to $0$ an entry $A_{i,j}$). This emulates the closing of a door through a passage that for instance leads to a dead end or that does not bring efficiently the walker to the exit.
    \item[2 --] Breaking walls. The system can change the adjacency matrix of the environment adding a link (changing from $0$ to $1$ an entry $A_{i,j}$). This mimics the creation of an hole in a maze wall, hence opening a shortcut.
\end{itemize}

In this setup, the objective function is the amount of time required to exit the maze (to be minimized) or the amount of population that exits to the sink in a given amount of time (to be maximized). If the adjacency matrix can change -- either intrinsically by random flips or as a result of actions taken by the agent -- we are in the canonical scenario for RL but in a quantum domain (quantum RL). We thus want to learn a policy by which, dynamically changing the topology of the maze, even by just adding or removing a small number of walls, we can  significantly improve the transfer rate of the  walker to the sink. We can also imagine the action of modifying the adjacency matrix as a sort of additional, engineered {\it noise} that, if carefully tuned, can improve the exit performance of the stochastic quantum walker.
An example of such an improvement is shown in Fig.\ref{fig:training_curve} where an agent was trained to perform the described actions on a $6 \times 6$ perfect maze. We can see how the performance improves from the baseline stochastic quantum walker as the agent learns how to get better rewards, i.e. transferring more population into the sink. The number of actions made by the agent was quite small ($8$ in this particular case) but carefully planned in order to yield a great boost to the efficiency of the walker. Given the key role of transport of energy or information over complex networks, this approach (see more details in Ref. \cite{maze2020preparation}) might represent an interesting and promising step towards novel QML schemes for NISQ devices and quantum technologies in general.

\section{Conclusions}

Quantum Machine Learning is where nowadays machine learning is going to meet quantum information science in order to realize more powerful quantum technologies. However, a much deeper understanding of their underlying mechanisms is still required in order to develop new algorithms and especially to apply them to address real problems and then to lead to commercial applications. This is a very young but very rapidly developing research field where the the first results pave the way for new experimental demonstrations of such hybrid classical-quantum protocols allowing to evaluate the potential advantages of exploiting them over their classical ML counterparts and then to exploit them on commercially available or coming soon NISQ devices and quantum technologies. We have hereby presented a perspective on new recent algorithms covering the main areas of ML (supervised, unsupervised and reinforcement learning) and different combinations of quantum-classical data and algorithms (QQ, QC, CQ models). We have also pointed out that there is still plenty of work to do, as it is definitely necessary to scale up the algorithms listed here (and all the other models that have been proposed recently) to the limits of existing devices performing an accurate scaling analysis of performances and corresponding errors. Ultimately, the crucial question we will need to answer in this field is whether a quantum speedup is theoretically and experimentally feasible via Quantum Machine Learning models running on NISQ machines.
 
\bibliographystyle{unsrt}
\bibliography{bibliography}

\end{document}